\documentclass[aps,showpacs,prl,twocolumn,floatfix]{revtex4}
\usepackage{epsfig,amssymb,amsfonts}

\makeatletter
\renewcommand{\@makefntext}[1]{\parindent=0em\noindent\hbox to 0.9em{\hss$^{\@thefnmark}$}#1}
\renewcommand{\@footnotemark}{\hbox{\mathsurround=0pt$^{\@thefnmark}$}}
\newcommand{\ftnote}[2]{\footnotemark[#1]\footnotetext[#1]{#2}}
\makeatother
\DeclareMathSymbol{\varGamma}{\mathord}{letters}{"00}
\begin{document}

\title{Comment on ``Once more about the $K\bar{K}$ molecule approach to the light scalars''}

\author{Yu. S. Kalashnikova}
\affiliation{Institute of Theoretical and Experimental Physics, 117218, B.Cheremushkinskaya 25, Moscow, Russia}

\author{A. E. Kudryavtsev}
\affiliation{Institute of Theoretical and Experimental Physics, 117218, B.Cheremushkinskaya 25, Moscow, Russia}

\author{A. V. Nefediev}
\affiliation{Institute of Theoretical and Experimental Physics, 117218, B.Cheremushkinskaya 25, Moscow, Russia}

\author{J. Haidenbauer}
\affiliation{Institut f\"{u}r Kernphysik, Forschungszentrum J\"{u}lich GmbH, D--52425 J\"{u}lich, Germany}

\author{C. Hanhart}
\affiliation{Institut f\"{u}r Kernphysik, Forschungszentrum J\"{u}lich GmbH, D--52425 J\"{u}lich, Germany}

\newcommand{\vk}{{\vec k}}
\newcommand{\vq}{{\vec q}}
\def\No{{\char'374}}

\begin{abstract}
In this manuscript we comment on the criticism raised recently by Achasov and Kiselev [Phys. Rev. D 76, 077501 (2007)]
on our work on the radiative decays $\phi \to \gamma a_0/f_0$ [Eur. Phys. J. A 24, 437 (2005)].
Specifically, we demonstrate that their criticism relies on results that violate gauge--invariance and is therefore invalid.
\end{abstract}
\maketitle

In a recent paper \cite{mol} we considered the radiative decay
$\phi \to \gamma a_0/f_0$ in the molecular ($K\bar{K}$) model of the
scalar mesons ($a_0(980)$, $f_0(980)$). In particular, we showed that
there was no considerable suppression of the decay amplitude due to the
molecular nature of the scalar mesons. In addition, as a more general
result we demonstrated that, as soon as the vertex function of the
scalar meson is treated properly, the corresponding loop integrals
become very similar to those for point--like (quarkonia) scalar mesons,
provided reasonable values are chosen for the range of the interaction.
We also confirmed the range of order of $10^{-3}\div 10^{-4}$
for the branching ratio obtained in Refs.~\cite{Oset,Markushin,Oller}
within the molecular model.

As a reaction to our work a paper appeared \cite{AK}, where the
authors criticize our results and claim that our paper \cite{mol}
is ``misleading''. Specifically, they dispute our findings that the
transition amplitude $\phi \to K^+K^- \to \gamma a_0/f_0$ is governed
by low kaon-momenta (nonrelativistic kaons) in the loop. In order to
support this conjecture they present numerical results that supposedly
demonstrate that ``ultrarelativistic kaons determine the real part of
the $\phi \to K^+K^- \to \gamma a_0/f_0$ amplitude''. The dominance
of such contributions of ``kaon high virtualities'' is then interpreted
as support for a compact four-quark nature of the scalar mesons.

In this comment we want to point out a fundamental flaw in the
calculations presented in Ref. \cite{AK} which, in turn, invalidates
the criticism raised in that paper. Namely, in order to
demonstrate that the high-momentum components determine the
$\phi \to K^+K^- \to \gamma a_0/f_0$ amplitude the authors of \cite{AK}
introduce a momentum cut-off in the relevant integrals. However,
in doing so  gauge--invariance gets violated. As will be shown
below, large momentum contributions appear only in this induced
gauge--invariance--violating term and are therefore of no physical significance.

\begin{figure}[t]
\begin{center}
\begin{tabular}{cc}
\epsfig{file=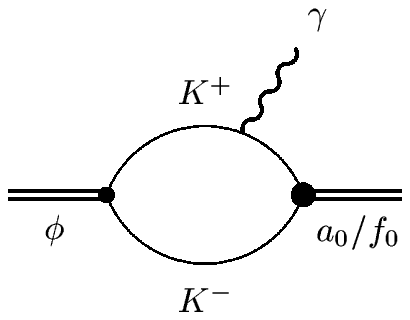,width=3.5cm}&\raisebox{-6mm}{$\epsfig{file=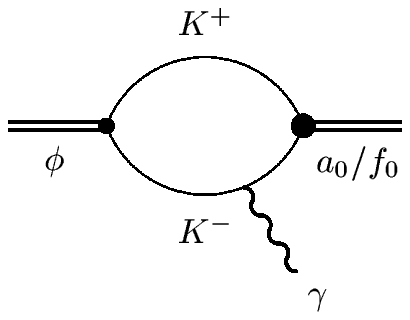,width=3.5cm}$}\\
(a)&(b)\\
\epsfig{file=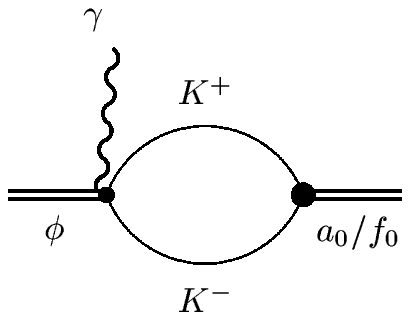,width=3.5cm}&\epsfig{file=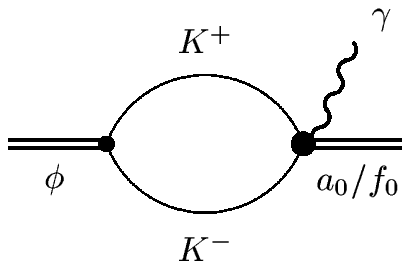,width=3.5cm}\\
(c)&(d)
\end{tabular}
\end{center}
\caption{Diagrams contributing to the amplitude of the radiative decay $\phi \to \gamma a_0/f_0$.}
\label{diags}
\end{figure}

To keep our argument self contained we briefly repeat the essentials
of the formalism. As a consequence of gauge--invariance the full
matrix element for the $\phi\to\gamma S$ ($S$ = $a_0$ or $f_0$) decay,
$M_\nu$,
can be written as
\begin{eqnarray}
\nonumber
M_{\nu}&=&\frac{eg_{\phi}g_S}{2\pi^2im^2}I(m_V,m_S)
[\varepsilon_{\nu}(p\cdot q)-p_{\nu} (q\cdot \varepsilon)]\\
&=&eg_{\phi}g_S\varepsilon^{\mu}J_{\mu\nu},
\label{loop}
\end{eqnarray}
where $p$ and $q$ are the momenta of the $\phi$ meson and
the photon, respectively, $m$ is the kaon mass,
$g_\phi$ and $g_S$ are the $\phi K^+K^-$ and $SK^+K^-$
coupling constants, and
$\varepsilon_\nu$ is the polarization four--vector of the
$\phi$ meson. The masses of the $\phi$ meson and the scalar are denoted
by $m_V$ and $m_S$, respectively.
The function $I(m_V,m_S)$  has a smooth limit for $q\to 0$.
As a consequence of gauge--invariance
the amplitude (\ref{loop}) is transverse, $M_{\nu}q^{\nu}=0$,
and is proportional to the photon momentum; especially
it vanishes for $q\to 0$. The form
(\ref{loop}) is well known. Details can be found, for
example, in Refs. \cite{AI,Nussinov,Lucio,Bramon,CIK}.

For point--like scalars, only diagrams (a)--(c) of Fig. \ref{diags}
contribute. If the scalars are regarded as extended objects, a vertex
function needs to be introduced at the $\bar KKS$ vertex.  Then gauge
invariance demands the inclusion of a diagram of type (d). For general
kinematics a proper construction of this additional term is quite
involved (see Ref.~\cite{BS} where we list a few of the papers
devoted to this subject) and contains some ambiguity. However,
for soft photons, all the different recipes give the same
result up to corrections of order $(q/\beta)^2$ that will be dropped.
Here $1/\beta$ denotes the range of forces --- for the case of interest
one may use $\beta\sim m_\rho$, where $m_\rho$ denotes the mass of the
lightest exchange particle allowed, namely that of the $\rho$ meson~\cite{mol}.
We may then as well  use the method suggested in Ref.~\cite{CIK}
that is based on minimal substitution
considerations. For more details on the issue of gauge invariance
for the reaction considered here, see Ref.~\cite{firstversion}.

After this introduction let us discuss the main
formula of our paper \cite{mol}. It is argued there (and
confirmed by actual calculations) that, since the
amplitude is finite even for the point--like limit,
the range of convergence of the integrals involved
{\em is defined only by the kinematics of the problem}. In
particular, if both masses, i.e. that of
the vector and of the scalar meson, are close to the
$K \bar K$ threshold, the integrals
converge at $k_0\sim m$,  and thus for non--relativistic values of
the three--dimensional loop momentum $\vk$, $|\vk| \ll m$.
This allows us to perform a nonrelativistic reduction of the
amplitude in the rest-frame of the $\phi$ meson.
The integrals in question for the individual graphs of Fig.~1
are (note that $J^{(b)}_{ik}=J^{(a)}_{ik}$):
\begin{eqnarray}
J^{(a)}_{ik}&=&\frac{-i}{2m^3}\int\frac{d^3 k}{(2\pi)^3}
\frac{k_iK_k\varGamma(K)}{[E_V-\frac{k^2}{m}+i0][E_S-\frac{K^2}{m}+i0]},\nonumber \\
J^{(c)}_{ik}&=&\frac{-i}{2m^2}\delta_{ik}\int\frac{d^3 k}{(2\pi)^3}
\frac{\varGamma(k)}{E_S-\frac{k^2}{m}+i0},\label{explint}\\
J^{(d)}_{ik}&=&\frac{-i}{2m^2}\int \frac{d^3k}{(2\pi)^3}
\frac{k_ik_k}{E_V-\frac{k^2}{m}+i0}\frac{1}{k}\frac{\partial \varGamma(k)}{\partial k},\nonumber
\end{eqnarray}
where $E_V=m_V-2m$, $E_S=m_S-2m$, and $\vec K =\vk-\frac{1}{2}\vq$. The
last integral can be rewritten by performing an integration by parts:
\begin{eqnarray}
\nonumber
 J^{(d)}_{ik}&=& \frac{i}{2m^2}\delta_{ik} \int
\frac{d^3 k}{(2\pi)^3}
\frac{\varGamma(k)}{E_V-\frac{k^2}{m}+i0} \\
\nonumber
&+&
\frac{i}{3m^3}\delta_{ik} \int \frac{d^3 k}{(2\pi)^3}
\frac{k^2\varGamma(k)}{[E_V-\frac{k^2}{m}+i0]^2}\\
&-&\frac{i}{12\pi^2m^2}\delta_{ik} \int_0^\infty d k
\frac{\partial}{\partial k}\left(
\frac{k^3\varGamma(k)}{E_V-\frac{k^2}{m}+i0}\right).
\label{nbyparts}
\end{eqnarray}
Here, contrary to Ref.~\cite{mol}, with the last term we kept the surface
integral that emerges in the calculation.
In order to investigate the range of momenta relevant for the loop
integrals in Ref.~\cite{AK}, a momentum cut-off $\Lambda$ was introduced.
We follow this prescription and write the
full transition current as:
\begin{equation}
J_{ik}(\Lambda) = \hat J_{ik}(\Lambda) + \delta_{ik}R  (\Lambda) \, ,
\label{fullanswer}
\end{equation}
where
\begin{equation}
R(\Lambda) = -\frac{i}{12\pi^2m^2}
\frac{\Lambda^3\varGamma(\Lambda)}{E_V-\frac{\Lambda^2}{m}+i0}\nonumber\
\end{equation}
contains the above mentioned surface term
and
\begin{widetext}
\begin{eqnarray}
\hat J_{ik}(\Lambda) &=&-\frac{i}{m^3}\int^{\Lambda}\frac{d^3 k}{(2\pi)^3}\left\{
\frac{k_i(\vk-\frac{1}{2}\vq)_k\varGamma(\vk-\frac{1}{2}\vq)}{[E_V-\frac{k^2}{m}+i0][E_V-q
-\frac{(\vk-\frac12\vq)^2}{m}+i0]}\right.
\nonumber\\
&+&\left.\varGamma(k)\delta_{ik}\left(\frac{m}{2}\frac{q}{[E_V-q-\frac{k^2}{m}+i0][E_V-\frac{k^2}{m}+i0]}
-\frac13\frac{k^2}{[E_V-\frac{k^2}{m}+i0]^2}
\right)\right\} \, .
\end{eqnarray}
\end{widetext}

For later convenience we used energy conservation to replace $E_S$ via
$E_S=E_V-q$\ftnote{1}{Contrary to the claim made in reference [5] of
Ref.~\cite{AK} energy and momentum conservation are maintained in
the calculations of Ref.~\cite{mol}.}. For $\Lambda\to \infty$ $\hat
J_{ik}(\Lambda)$ matches to the formula used in Ref.~\cite{mol} to
calculate the matrix element for $\phi \to \gamma a_0/f_0$. We checked
that this sum of integrals converges for non--relativistic kaon momenta.
This finding was confirmed in Ref.~\cite{AK}.

\begin{figure}[t]
\begin{center}
\epsfig{file=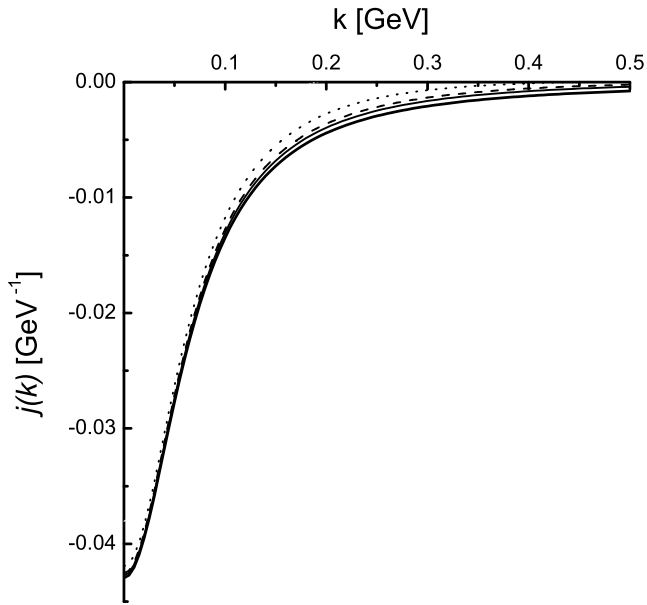,width=8.5cm}
\caption{The behaviour of the integrand $j(k)$
of Im$(\hat J_{ik})$, as a function of the kaon momentum floating in the
loop. We chose $m_S=0.98$ GeV, $m_V=1.02$ GeV,
$m=0.495$ GeV, and  four values of the parameter $\beta$:
$\beta=0.4$ GeV (dotted line), $\beta=0.6$ GeV (dashed line),
$\beta=0.8$ GeV (thin solid line), and $\beta=\infty$ (thick solid
line).}\label{imj}
\end{center}
\end{figure}

For illustration we choose a particular form of $\varGamma(k)$, namely
$\varGamma(k)=\beta^2/(k^2+\beta^2)$, and study that part of $\hat J_{ik}$
proportional to the structure $\delta_{ik}$ (according to Eq.~(\ref{loop})
exactly this structure contributes to the decay amplitude in the $\phi$-meson
rest frame). In Fig.~\ref{imj} we plot the behaviour of the integrand $j(k)$ (
$Im(\hat J_{ik})=\delta_{ik}\int_0^\infty\;j(k)\;dk$),
as a function of $k$
(note, that the
integrand in the similar integral for Re$(\hat J_{ik})$ contains
$\delta(k^2-mE_V)$, such that $k=\sqrt{mE_V}\approx 0.12$ GeV). From
Fig.~\ref{imj} one can see that the integral indeed converges at
non--relativistic values of the kaon momentum, regardless of the value
of the finite-range parameter $\beta$ --- the latter plays no role
for the convergence.

In Ref.~\cite{mol} the last term of Eq.~(\ref{fullanswer}), $R(\Lambda)$, was
dropped, for it vanishes exactly for $\Lambda\to \infty$\ftnote{2}{For
this to be true we only need to demand that $\lim_{\Lambda\to
\infty}\Lambda \Gamma(\Lambda)=0$.}. In Ref.~\cite{AK}, however, it
is argued that this term should be kept and that
it converges only for very large values of $\Lambda$, which
means that the corresponding integral
acquired
contributions from very large momenta. The contribution of those large
momentum components is then taken as a proof that only if the scalars
are very compact objects, a sizable contribution from the loop can
emerge. Notice that, even for finite values of $\Lambda$,
$\hat J_{ik}(\Lambda)$ vanishes for $q\to 0$,
as required by the general structure given in Eq.~(\ref{loop}).
However, since $R(\Lambda)$ is independent of the photon momentum $q$,
it gives a nonvanishing contribution to $J_{ik}$ even for $q=0$ for
all finite values of $\Lambda$.  Therefore this term violates
gauge--invariance. Thus, by introducing a sharp cut-off into the
problem the authors of Ref.~\cite{AK} produced a term that violates
gauge--invariance\ftnote{3}{With sharp cut-off, gauge--invariance of
the amplitude can be restored by a subtraction at $q=0$. Obviously,
this procedure is equivalent to omission of the last,
$q$--independent term in Eq.~(\ref{fullanswer}).}. Since the whole
argument presented in Ref.~\cite{AK} is based on this term, it bears
no physical significance.

We therefore conclude that all results of Ref.~\cite{mol} are
valid. In particular, there is no strong suppression of kaon loops by
the scalar wave function. Regardless of this,
it should be stressed that the data for $\phi\to \gamma a_0/f_0$~\cite{data} is very sensitive
to the nature of the light scalar mesons, for it allows direct
access to the effective coupling constant $g_{\rm eff}$ of the scalar
to the kaons. As was shown in Ref.~\cite{evidence} this coupling is a
direct measure of the molecular contribution of the scalar mesons.

As a reply to this comment, the authors of the commented paper \cite{AK} state that, with a properly regularised 
amplitude $T_{\nu\mu}$ (up to the overall normalisation $T_{\nu\mu}$ coinsides with our $J_{\nu\mu}$ introduced in 
Eq.~(\ref{loop}) above), {``the regulator field contribution is caused
fully by high momenta ($M\to\infty $) and teaches us how to allow
for high $K$ virtualities in gauge invariant way"} \cite{AK2}. To exemplify their statement, the authors of Ref.~\cite{AK2}
employ the Pauli-Villars regularisation in the full relativistic expression for the amplitude,
\begin{equation}
\overline{T}\left \{\phi(p)\to\gamma [a_0(q)/f_0(q)],M\right\}=\epsilon^\nu(\phi)\epsilon^\mu(\gamma)\overline{T}_{\nu\mu}(p,q),
\end{equation}
where the overline marks the regularised quantities. In particular, for a quantity
\begin{equation}
{\cal O}=\int d^4 r f(r,p;m),
\end{equation}
the corresponding regularised integral reads:
\begin{equation}
\overline{\cal O}(M)=\int d^4 r [f(r,p;m)-f(r,p;M)],
\end{equation}
with $M$ being the regulator mass; the physical amplitude corresponds to the limit $M\to\infty$. Then, by an
explicit calculation, the authors of Ref.~\cite{AK2} find that
\begin{eqnarray}
\nonumber
\epsilon^\nu(\phi)\epsilon^\mu(\gamma)\overline{T}_{\nu\mu}(p,p) \\
& &\hspace{-2.5cm}=\epsilon^\nu(\phi)\epsilon^\mu(\gamma)\left(T^{m}_{\nu\mu}(p,p)
-T^M_{\nu\mu}(p,p)\right) \nonumber\\
& &=(\epsilon(\phi)\epsilon(\gamma))(1-1)=0,
\end{eqnarray}
and thus they conclude that the regularised amplitude is gauge invariant, and this is due to the final subraction
coming from the regulator piece.  
To have a deeper insight, consider $T_{\nu\mu}^m(p,p)$ and extract the coefficient of the structure $g_{\mu\nu}$. 
Then, after using the Feynman parametisation, one has:
\begin{eqnarray}
T^{m}_{\nu\mu}(p,p)=C\left[g_{\mu\nu}\int^1_0dz\int\frac{d^4r}{(2\pi)^4}\frac{1}{(r^2-R^2)^2}\right.\nonumber\\
-8\int^1_0 dz(1-z)\int\left.\frac{d^4r}{(2\pi)^4}\frac{r_{\mu}r_{\nu}}{(r^2-R^2)^3}\right],\label{tpp}
\end{eqnarray}
where $C$ is an unimportant numerical coefficient and $R^2=z(1-z)p^2-m^2$. The integrand can be rewritten in the form:
\begin{equation}
\frac{r^2g_{\mu\nu}-4r_{\mu}r_{\nu}}{(r^2-R^2)^3}-\frac{g_{\mu\nu}R^2}{(r^2-R^2)^3}.
\label{int1}
\end{equation}
It is tempting now to perform averaging over the angular variables
first, substituting $r_{\mu}r_{\nu}\to\frac14 r^2 g_{\mu\nu}$. Then,
naively, the first term in the integrand (\ref{int1}) vanishes,
whereas the remaining, second term gives a finite constant independent
of the mass $m$.  Conclusions made by the authors of Ref.~\cite{AK2}
are based on this result, and, finally, they notice that ``the
finiteness of the subtraction constant hides its high momentum
origin".  However, the analysis performed above has a flaw.  Indeed,
although the angular integration of the first term in (\ref{int1})
gives zero, the remaining radial integral diverges.  Therefore, one
deals with an undefined expression of the kind $0\times\infty$.  In
order to resolve this issue, it is important to deal with finite
integrals. To this end, we evaluate the integral in $d=4-\epsilon$
dimensions. The
radial integral is finite now, whereas the angular integral gives the
substitution $r_{\mu}r_{\nu}\to\frac{1}{4-\epsilon} r^2 g_{\mu\nu}$.
It is easy to check that, after taking the limit $\epsilon\to 0$,
the contribution of the first term does not vanish any more but, on
the contrary, it cancels the contribution of the second term in
(\ref{int1}). Thus
\begin{equation}
T^m_{\nu\mu}(p,p)=0\quad\mbox{and}\quad T^M_{\nu\mu}(p,p)=0
\end{equation}
individually.
Thus, the apparent contribution of high kaon virtualities found
in Ref.~\cite{AK2} appears as a result of misusing of 
the Pauli--Villars regularisation scheme, when a regularised finite integral is artificially split into
divergent parts and each part is considered separately. In other words, the physical 
amplitude is given by a sum of several divergent integrals and, in any 
correct regularisation scheme, these infinite parts cancel each other 
exactly. 

Finally, had the infinitely high momenta been relevant, as advocated in Refs.~\cite{AK,AK2}, 
then the pointlike limit would have rendered useless, as there is no infinitely compact states in nature. 
This becomes clear when a scalar form factor is introduced, in 
which case neither dimensional nor Pauli--Villars regularisation is needed.
In order to maintain gauge invariance one is forced either to include the 
diagram (d) of Fig.~1, or to subtract explicitly the 
amplitude, given by the sum of the diagrams (a)-(c), at $q \to 0$, 
with both procedures being equivalent. Notice 
that gauge invariance requires the subtraction of the original amplitude, 
and not the one with mass $m$ replaced by the (infinitely large) 
regulator mass $M$. The latter observation invalidates the 
claim made in Ref.~\cite{AK2}.

\begin{acknowledgments}
The authors acknowledge useful discussions with N. N. Acha\-sov and
A. Kiselev. This research was supported by the grants
RFFI-05-02-04012-NNIOa, DFG-436 RUS 113/820/0-1(R), NSh-843.2006.2,
and NSh-5603.2006.2, by the Federal Programme of the Russian
Ministry of Industry, Science, and Technology 40.052. 1.1.1112, and
by the Russi\-an Go\-vern\-men\-tal Agreement 02.434.11.7091. A.E.K.
acknowledges also partial support by the grant DFG-436 RUS 113/733.
A.N. would also like to acknowledge the financial support via the project
PTDC/FIS/70843/2006-Fisica and of the non-profit ``Dynasty"
foundation and ICFPM.  
\end{acknowledgments}

\end{document}